\documentclass{nature}
\spacing{1}

\PassOptionsToPackage{warn}{textcomp}
\usepackage{times}
\usepackage[ngerman,english]{babel}
\usepackage{upgreek}
\usepackage{units}
\usepackage{graphicx}
\usepackage{amsmath}
\usepackage{amssymb}
\usepackage{amsthm}
\usepackage{mathcomp}
\usepackage{mathrsfs}
\usepackage{braket}
\usepackage{dsfont}
\usepackage{esvect}
\usepackage{esdiff}

\usepackage{float}
\usepackage{url}
\usepackage{framed}
\usepackage{xcolor, xspace}
\usepackage {multicol}
\usepackage{ulem}
\usepackage[displaymath,mathlines]{lineno}

\newcommand*\patchAmsMathEnvironmentForLineno[1]{%
  \expandafter\let\csname old#1\expandafter\endcsname\csname #1\endcsname
  \expandafter\let\csname oldend#1\expandafter\endcsname\csname end#1\endcsname
  \renewenvironment{#1}%
     {\linenomath\csname old#1\endcsname}%
     {\csname oldend#1\endcsname\endlinenomath}}%
\newcommand*\patchBothAmsMathEnvironmentsForLineno[1]{%
  \patchAmsMathEnvironmentForLineno{#1}%
  \patchAmsMathEnvironmentForLineno{#1*}}%
\AtBeginDocument{%
\patchBothAmsMathEnvironmentsForLineno{equation}%
\patchBothAmsMathEnvironmentsForLineno{align}%
\patchBothAmsMathEnvironmentsForLineno{flalign}%
\patchBothAmsMathEnvironmentsForLineno{alignat}%
\patchBothAmsMathEnvironmentsForLineno{gather}%
\patchBothAmsMathEnvironmentsForLineno{multline}%
}

\usepackage{multibib}
\newcites{sec}{References Methods}

\newcommand*\downk{\ket{\downarrow}}

\newcommand*\downkmg{\ket{\downarrow}_\mathrm{Mg}}

\newcommand*\upk{\ket{\uparrow}}

\newcommand*\upkmg{\ket{\uparrow}_\mathrm{Mg}}

\newcommand*\half{\frac{1}{2}}
\newcommand*\ii{\mathrm{i}}
\newcommand*\dprime{{\prime\prime}}
\newcommand*\dif{\mathrm{d}}

\def\fm#1{\ifmmode #1 \else $#1$\fi}
\def\MgH{\fm{{}^{24}\mathrm{MgH}^{+}}\xspace}
\def\Mgtf{\fm{{}^{25}\mathrm{Mg}^{+}}\xspace}
\def\DMgH{\fm{\Delta_{\mathrm{MgH}}}\xspace}
\def\DMg{\fm{\Delta_{\mathrm{Mg}}}\xspace}
\def\RD{\fm{R(\DMgH)}\xspace}
\def\upkm{\fm{\upk_\mathrm{m}}\xspace}
\def\downkm{\fm{\downk_\mathrm{m}}\xspace}
\def\molres{\fm{\mathrm{X}^1\Sigma^+(J=1)\rightarrow\mathrm{A}^1\Sigma^+(J=0)}\xspace}

\title{Non-destructive state detection for quantum logic spectroscopy of molecular ions}

\author{Fabian Wolf$^{1}$, Yong Wan$^{1\dagger}$, Jan C. Heip$^1$, Florian Gebert$^1$, Chunyan Shi$^1$ \& Piet O. Schmidt$^{1,2}$ }

\begin{document}

\maketitle

\begin{affiliations}
 \item Physikalisch-Technische Bundesanstalt, 38116 Braunschweig, Germany
 \item Institut f\"ur Quantenoptik, Leibniz Universit\"at Hannover, 30167 Hannover, Germany
 \item Present address: National Institute of Standards and Technology, 325 Broadway, Boulder, CO 80305, USA
\end{affiliations}
\begin{abstract}

Precision laser spectroscopy\cite{germann_observation_2014} of cold and trapped molecular ions is a powerful tool for fundamental physics, including the determination of fundamental constants\cite{koelemeij_vibrational_2007}, the laboratory test for their possible variation\cite{schiller_tests_2005,flambaum_enhanced_2007}, and the search for a possible electric dipole moment of the electron\cite{loh_precision_2013}.
While the complexity of molecular structure facilitates these applications, the absence of cycling transitions poses a challenge for direct laser cooling\cite{shuman_laser_2010}, quantum state control\cite{rellergert_evidence_2013, schneider_all-optical_2010, staanum_rotational_2010, lien_broadband_2014, hansen_efficient_2014}, and detection.
Previously employed state detection techniques based on photo-dissociation\cite{hojbjerre_rotational_2009} or chemical reactions\cite{tong_sympathetic_2010} are destructive and therefore inefficient, restricting the achievable resolution in laser spectroscopy.
Here we experimentally demonstrate non-destructive state detection of a single trapped molecular ion through its strong Coulomb coupling to a well-controlled co-trapped atomic ion.
An algorithm based on a state-dependent optical dipole force\cite{leibfried_experimental_2003} (ODF) changes the internal state of the atom conditioned on the internal state of the molecule.
We show that individual quantum states in the molecular ion can be distinguished by their coupling strength to the ODF and observe black-body radiation-induced quantum jumps between rotational states of a single molecular ion. Using the detuning dependence of the state detection signal, we implement a variant of quantum logic spectroscopy\cite{schmidt_spectroscopy_2005, wan_precision_2014} of a molecular resonance.
The state detection technique we demonstrate is applicable to a wide range of molecular ions, enabling further applications in state-controlled quantum chemistry\cite{ratschbacher_controlling_2012} and spectroscopic investigations of molecules serving as probes for interstellar clouds\cite{brunken_h2d+_2014,campbell_laboratory_2015}.
\end{abstract}
One of the salient features of trapped ion systems is that the universal Coulomb interaction allows strong coupling of diverse quantum objects, such as different species of atomic ions or atomic and molecular ions. Being able to perform quantum logic operations e.g. in the form of gates\cite{cirac_quantum_1995,schmidt-kaler_realization_2003,leibfried_experimental_2003} between the quantum objects has proven a powerful tool for quantum information processing and quantum simulations in such systems.
It also allows combining the advantages of different atomic species. Quantum logic spectroscopy is one such application in which the high degree of control achieved over selected atomic ions is extended to species over which such control is lacking\cite{schmidt_spectroscopy_2005, wan_precision_2014}. Here, we demonstrate for the first time quantum logic operations between a single molecular ion and a co-trapped atomic ion, making a wide range of molecular ions accessible to this highly-developed toolbox. The presented technique allows the investigation of single molecules in a well isolated system avoiding disturbance from the environment, which is the limiting factor in other implementations of single molecule spectroscopy such as surface enhanced Raman spectroscopy (SERS)\cite{nie_probing_1997}.

Quantum logic operations between atoms are based on state dependent forces often induced by laser fields.
The same approach is applicable to molecular ions. The coupling is now distributed over many ro-vibrational transitions, which weakens the interaction strength for individual transitions (see Methods). The energy splitting between quantum states allows the ODF to distinguish between them, facilating rotational state detection.
By tuning the frequency of a laser field near one of the molecular resonances, an ODF is exerted on the molecular ion if it is in one of the states coupled by the corresponding transition.
If the strength of this force is oscillating with a frequency close to one of the secular motional frequencies of the two-ion crystal in the trap, it is resonantly enhanced, generating coherent states of motion, which can be detected on a co-trapped atomic ion\cite{schmidt_spectroscopy_2006, vogelius_probabilistic_2006, hume_trapped-ion_2011}.
\begin{figure}[htb]
\centering
\includegraphics[width=89mm]{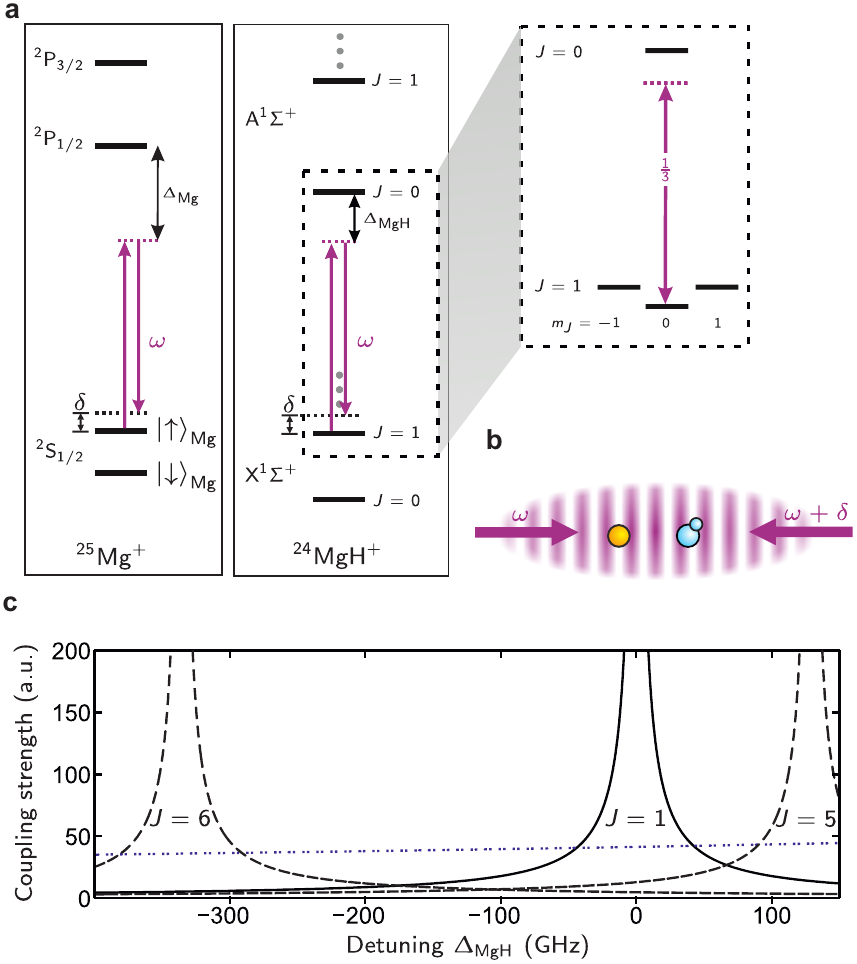}
\caption{\textbf{Coupling strength of an optical dipole force to the atomic and molecular ions.}
\textsf{(a)} Partial level scheme of the relevant atomic and molecular states and their laser coupling to the lattice laser. The inset shows the molecular substructure for the states involved in the sequence. The number on the transition is the experimentally implemented geometric coupling coefficient.
\textsf{(b)} The two lattice beams form a moving interference pattern leading to a temporally and spatially varying light force on both ions.
\textsf{(c)} Rotational state ($J,m_J=0$) and detuning-dependent coupling strength of the dipole force interacting with a \MgH molecular ion on the P-branch~($J\rightarrow J-1$, solid line) and R-branch~($J\rightarrow J+1$, dashed line) of the $\mathrm{X}^1\Sigma^+\rightarrow\mathrm{A}^1\Sigma^+$ electronic transition. The blue dotted line is the coupling strength to the \Mgtf ion.
}
\label{fig:couplingstrength}
\end{figure}
\begin{figure}
\centering
\includegraphics[width=89mm]{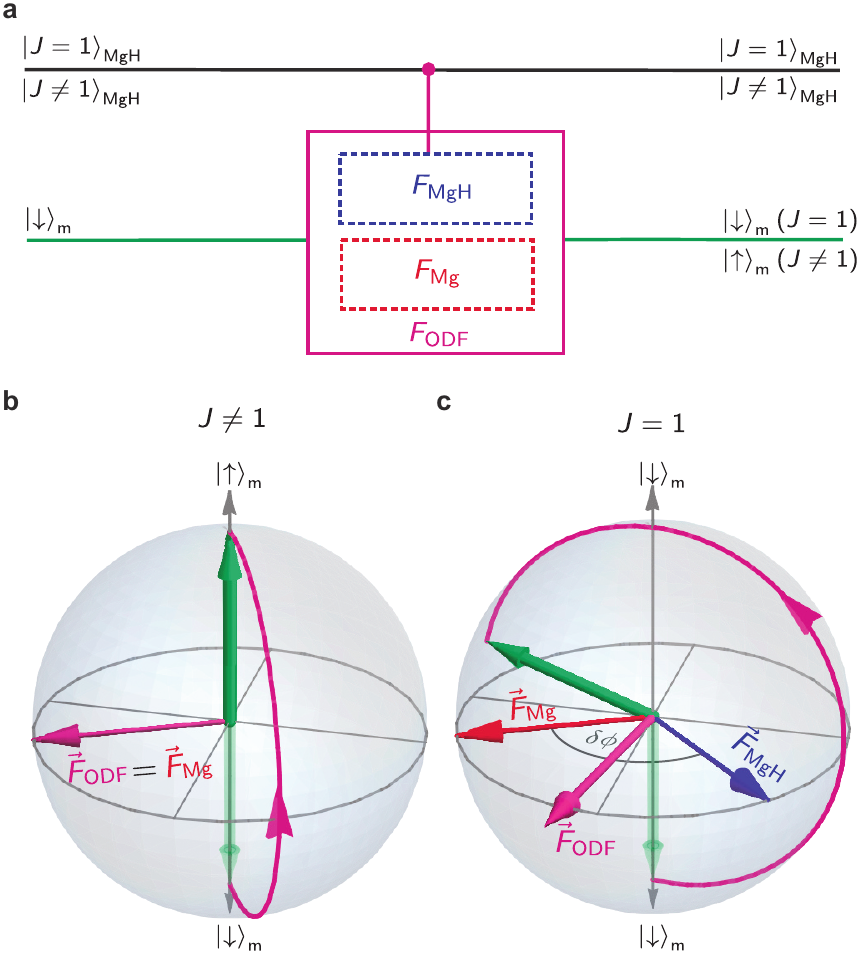}
\caption{\textbf{Molecular state mapping to the motional qubit.} (\textsf{a}) Circuit description of the mapping process.  For ideally chosen parameters $\DMgH$ and $\delta\phi$, as shown, the circuit is equivalent to a CNOT. The action of the ODF on the motional qubit is represented by the torque $\vec{F}_\mathrm{ODF}=\vec{F}_\mathrm{Mg}+\vec{F}_\mathrm{MgH}$. (\textsf{b}) For $J\neq1$ only the atom interacts with the lattice. The ODF creates a torque $\vec{F}_\mathrm{Mg}$ on the Bloch sphere (purple arrow). \textsf{(c)} If $J=1$, the additional force on the molecule (torque $\vec{F}_\mathrm{MgH}$ depicted in blue) is added to the force on the atom leading to a modified population transfer.
}
\label{fig:sequence}
\end{figure}

We extend the simple detection scheme described above to detect the small ODF arising from a laser field tuned near the \molres resonance (detuning \DMgH) in a \MgH molecular ion in the presence of a nearly detuning-independent background force from a co-trapped \Mgtf atomic ion (Fig.~\ref{fig:couplingstrength}\textsf{c}). Blackbody radiation (BBR) induces transitions between rotational states $J$, resulting in a thermal rotational state population with a maximum at $J=4$ for temperatures near \unit[300]{K}. We have chosen to probe the $J=1$ state, with a population of around $\unit[8]{\%}$, for its long average dwell time of around $\unit[3]{s}$ (see Extended Data Fig.~\ref{fig:MCSimu}). Vibrational population with vibrational quantum number $\nu>0$ in \MgH is negligible at room temperature owing to the large vibrational spacing\cite{balfour_rotational_1972}. Applying the ODF resonant with either the IP or OP mode implements a displacement operator resulting in coherent states of motion\cite{hume_trapped-ion_2011}, which are impractical for state discrimination due to their non-orthogonality. Instead, we map the molecule's internal state information to an engineered motional 2-level quantum system (motional qubit) that is coupled by the ODF. This approach allows us to suppress the background force from \Mgtf and enables advanced coherent population transfer schemes, such as composite pulses\cite{schmidt-kaler_realization_2003}.

The state detection algorithm starts with both axial motional modes [in-phase (IP) and out-of-phase (OP)] of the two-ion crystal consisting of a \Mgtf and a \MgH ion initialized in the motional ground state $\ket{0}_\mathrm{m}\equiv\ket{0}_\mathrm{IP}\ket{0}_\mathrm{OP}$ via the atomic ion\cite{wan_efficient_2015}. Two long-lived hyperfine states in \Mgtf serve as the atomic qubit denoted by $\downkmg, \upkmg$ (Fig.~\ref{fig:couplingstrength}\textsf{a}), initialized to $\downkmg$. The motional qubit is prepared in $\downk_\mathrm{m}\equiv\ket{1}_\mathrm{IP}\ket{0}_\mathrm{OP}$ by driving a $\pi$-pulse on the blue sideband transition (BSB) of the IP mode on the atomic ion while changing its internal state from $\downkmg$ to $\upkmg$.
The ODF is implemented by interfering two counterpropagating laser beams along the axial direction of the ion trap, forming an oscillating 1D optical lattice. By adjusting the relative detuning $\delta$ between the two beams to the difference between axial IP and OP mode motional frequencies, $\delta=\omega_\mathrm{OP}-\omega_\mathrm{IP}=\omega_\mathrm{IP}(\sqrt{3}-1)$, a Raman coupling involving only the motional states $\downk_\mathrm{m}\leftrightarrow \upk_\mathrm{m}\equiv\ket{0}_\mathrm{IP}\ket{1}_\mathrm{OP}$ is implemented, resulting in Rabi flopping between the two composite states (Fig.~\ref{fig:sequence} and Extended Data Fig.~\ref{fig:LatticeRabiFlops}). The interaction between the lattice laser field and the motional qubit is described by the Hamiltonian (see Methods)
\begin{equation}
H=\hbar\Omega_\mathrm{eff}\eta_\mathrm{IP}\eta_\mathrm{OP}\left(a_\mathrm{IP}a_\mathrm{OP}^{\dagger}+a_\mathrm{IP}^\dagger a_\mathrm{OP}\right),
\label{eq:Hamiltonian}
\end{equation}
with the creation operator $a_{\mathrm{IP}(\mathrm{OP})}^\dagger$ for the IP(OP) mode. The effective Rabi frequency is given by
\begin{equation}
\Omega_\mathrm{eff}=\Omega_{\mathrm{Mg}}\sqrt{1-2\RD\cos(\delta\phi)+\RD^2},
\end{equation}
where $\RD=\Omega_{\mathrm{MgH}}(\DMgH)/\Omega_{\mathrm{Mg}}$ is the ratio between the Rabi frequencies of the Raman couplings in the molecular and atomic ion, depending on the detuning \DMgH. The distance between the ions determines the relative phase $\delta\phi$ between the forces on the ions, and $\eta_\mathrm{IP,OP}$ are the Lamb-Dicke parameters for the OP and IP modes, respectively\cite{wan_efficient_2015}. In a Bloch sphere picture, the effect of the ODF on the atom and the molecule can be illustrated as the vector sum of two torques $\vec{F}_\mathrm{ODF}=\vec{F}_\mathrm{Mg}+\vec{F}_\mathrm{MgH}$ acting on the motional qubit (Fig.~\ref{fig:sequence}\textsf{b,c}).
In the experiment, the time $t_\mathrm{latt}$ in which the lattice laser interacts with the ions is adjusted such that for $\Omega_{\mathrm{MgH}}\simeq 0$ (i.e. the molecular ion is not in $J=1$) a $\pi$-pulse is performed on the motional qubit (Fig.~\ref{fig:sequence}\textsf{b}). If the molecule is in the probed state ($\Omega_{\mathrm{MgH}}\neq 0$), the additional force changes the population transfer (Fig.~\ref{fig:sequence}\textsf{c}).
This step transfers the internal information of the molecular state to the motional qubit. The second BSB $\pi$-pulse maps the motional state population (containing the molecule's internal state information) to the atomic qubit. The population in state $\downkmg$ follows
\begin{eqnarray}
P_{\downkmg}=\frac{1}{2}\left[1+\cos\left(\pi\sqrt{1-2\RD\cos(\delta\phi)+\RD^2}\right)\right]
\label{eq:signalstrength}
\end{eqnarray}
and is detected using state-dependent fluorescence on the \Mgtf ion, where $\downkmg$ fluoresces and $\upkmg$ does not. The full sequence is shown in Extended Data Fig.~\ref{fig:fullsequence}.

\begin{figure*}
\centering
\includegraphics[width=183mm]{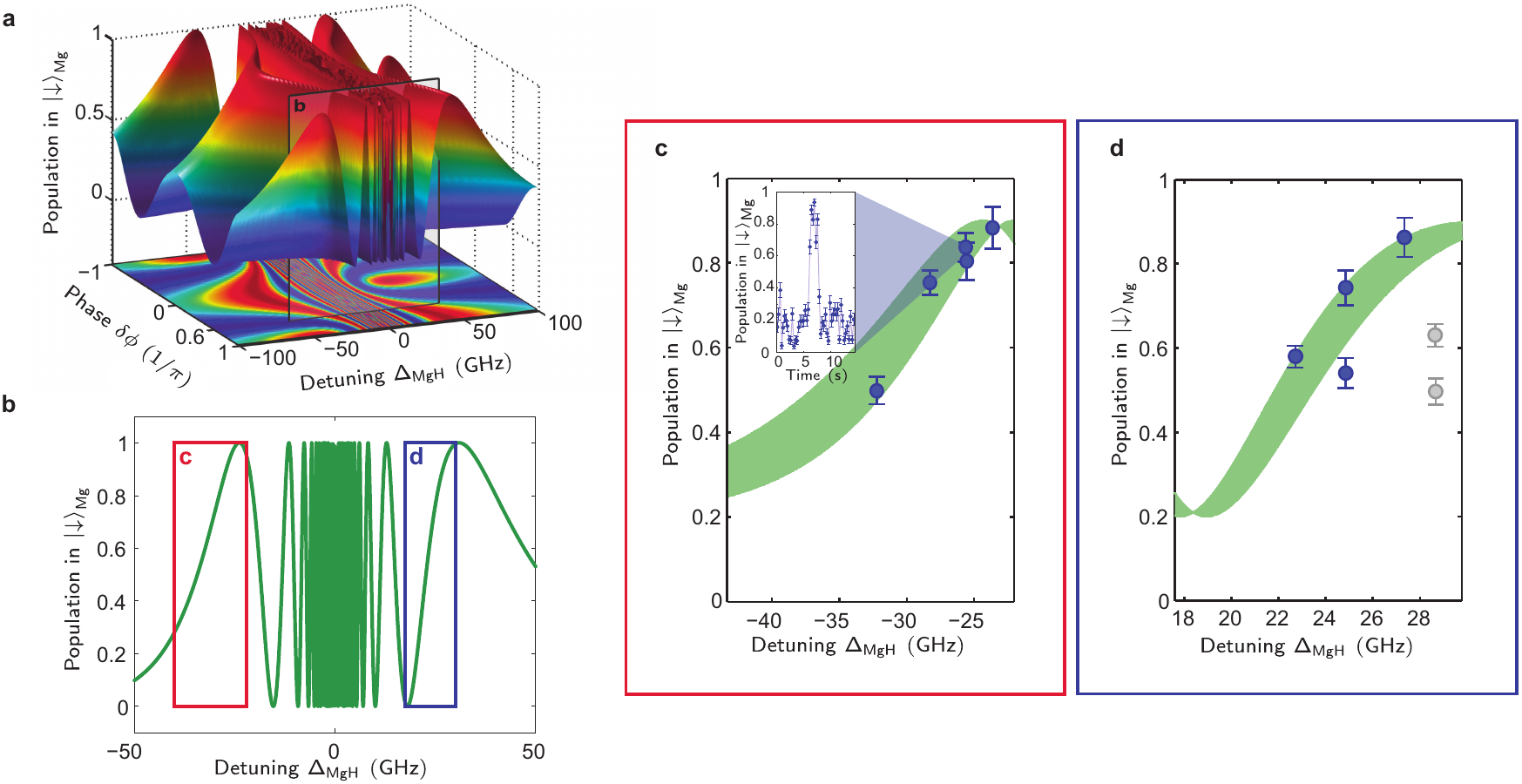}\footnotesize
\caption{\textbf{Non-destructive state detection.}
(\textsf{a}) Theoretically calculated signal $P_{\downkmg}$ on the atomic ion for \MgH in the $\ket{J=1,m_J=0}_\mathrm{MgH}$ state, and its dependence on the relative phase $\delta\phi$ of the optical force between the atomic and molecular ion and the detuning from molecular resonance.
(\textsf{b}) Cross section of \textsf{a}, for a phase $\delta \phi\approx 0.6\times \pi$, corresponding to a trap frequency of $\omega_\mathrm{IP}\approx \unit[2.17]{MHz}$.
(\textsf{c}) and (\textsf{d}) show the height of the observed signals as a function of the detuning for the regions indicated in \textsf{b}. A typical measurement result for a black-body radiation driven quantum jump into and out of the $J=1$ rotational state is shown in the inset. The error bars ($68~\%$ confidence interval) are dominated by quantum projection noise. The thickness of the theory bands accounts for small changes in trap frequency (and thus $\delta\phi$) for the different data points. The grey data points are excluded from the analysis in Fig.~\ref{fig:combineddata}, since they could not be assigned to a specific slope of the theoretical prediction, thus preventing unambiguous inversion of Eq. (\ref{eq:signalstrength}).
}
\label{fig:largefigure}
\end{figure*}

The calculated signal $P_{\downkmg}$ is shown in Fig.~\ref{fig:largefigure}\textsf{a} for different phases $\delta \phi$ and detunings $\DMgH$. The maxima correspond to rotations of $2 \pi n $ on the Bloch sphere of the motional qubit with $n\in\mathds{N}$. In particular, at the outermost peaks (see Fig.~\ref{fig:largefigure}\textsf{b}) the forces from the atomic and molecular ion cancel or add up to $2\pi$ exactly, depending on the sign of $\DMgH$.
There is a trade-off in the choice of lattice laser detuning from the molecular resonance: if the detuning is too large, the ODF will not be detectably different for a molecule in the rotational state of interest compared to a molecule in any other state. For a detuning too small, spontaneous Raman scattering will change the rotational state before it can be detected.
Under ideal experimental conditions, it is possible to choose a combination of the phase $\delta\phi$ and the detuning $\DMgH$ to perform state detection in a single experimental repetition (cycle), since it is free of quantum projection noise ($P_{\downkmg}=1$ with the molecules in $J=1$ and $P_{\downkmg}=0$ if it is not). In this case the experimental sequence can be understood in terms of a CNOT gate acting on the motional qubit with the molecular rotational qudit $\ket{J}_\mathrm{MgH}$ as a control (Fig.~\ref{fig:sequence}\textsf{a}):
\begin{align}
 \ket{J\neq 1}_\mathrm{MgH}\downkm &\longrightarrow\ket{J\neq 1}_\mathrm{MgH}\upkm,\\
 \ket{J\neq 1}_\mathrm{MgH}\upkm &\longrightarrow\ket{J\neq 1}_\mathrm{MgH}\downkm,\\
 \ket{J = 1}_\mathrm{MgH}\downkm &\longrightarrow\ket{J=1}_\mathrm{MgH}\downkm,\\
 \ket{J = 1}_\mathrm{MgH}\upkm &\longrightarrow\ket{J = 1}_\mathrm{MgH}\upkm.
\end{align}

A typical measurement with positive detection event can be seen in the inset of Fig.~\ref{fig:largefigure}\textsf{c}. If the molecular ion is not in the probed rotational state, the $\pi$-pulse on the motional qubit is successful and we observe low fluorescence (P$_{\downkmg}\sim 0$). After entering the $J=1$ state the additional force on the molecule will change the $\pi$-time on the motional qubit and we observe fluorescence on the \Mgtf ion (P$_{\downkmg}>0$), until the rotational state changes again. The signal corresponds to a BBR-induced quantum jump of the rotational state in the molecular ion. The signal background of P$_{\downkmg}\approx 0.2$ and reduced contrast arises from experimental imperfections in implementing the logic gates (see Methods). Since the algorithm implements a quantum non-demolition measurement of the molecular state, we reduce quantum projection noise by averaging each data point over 30 experimental cycles with a duration of $\sim 9.3$~ms per cycle.
An experimental event was identified as a positive detection event if at least three consecutive data points were significantly~(1.5$~\sigma$) above the background (see Methods).
In a total measuring time of approximately $\unit[23]{h}$ we have observed 18 positive detection signals of up to several seconds duration. The estimated average number of scattered photons from the optical lattice beams interacting with the molecular ion during a single detection cycle at $\DMgH=\unit[25]{GHz}$ is around 0.04 (see Methods).
If we assume that a single scattered photon removes the molecule from its rotational state, this results in an average single-cycle detection efficiency of $\unit[96]{\%}$, neglecting imperfections in any of the qubit and gate operations, thus representing an upper bound.
Estimates of the expected number of events using this detection efficiency together with the mean population and the required averaging time of $\sim\unit[1]{s}$ for a positive signal event agree well with the experimental results (see Methods).

The lattice light introduces an AC Stark shift on the order of several hundred $\unit{kHz}$, which is much larger than all magnetic couplings in the molecular ion. Therefore, the substates are labeled with $m_J$ in a basis with the quantization axis along the electric field direction of the lattice light. In this basis, only $\pi$-polarized light is present, coupling only to states with $m_J=0$ (see inset Fig.~\ref{fig:couplingstrength}\textsf{a}).
The resulting signal as a function of the detuning and molecular quantum state is shown in Fig.~\ref{fig:largefigure}\textsf{c} and \textsf{d} for $\delta \phi\approx 0.6\times \pi$, corresponding to a normal mode frequency of $\omega_\mathrm{IP}\approx \unit[2.17]{MHz}$. Experimental data for $\delta\phi\approx0$ is shown in Extended Data Fig.~\ref{fig:trapfreq2}.

\begin{figure}
\centering
\includegraphics[width=89mm]{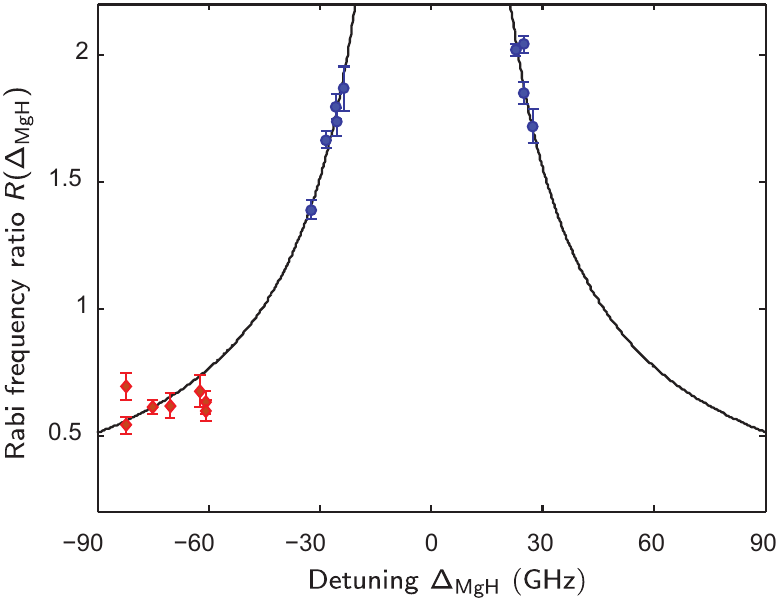}
\caption{\textbf{Quantum logic spectroscopy.} The Rabi frequency ratio $R= \Omega_\mathrm{MgH}(\DMgH)/\Omega_\mathrm{Mg}$ as a function of detuning $\DMgH$ from a molecular resonance for all data for which Eq.~\ref{eq:signalstrength} could be inverted unambiguously (see Fig.~\ref{fig:largefigure}\textsf{c},\textsf{d}).
Measurement data recorded at $\omega_\mathrm{IP}\approx \unit[2.17]{MHz}$ ($\delta \phi\approx 0.6\times\pi$, 9 Points, see Fig.~\ref{fig:largefigure}) and $\omega_\mathrm{IP}\approx \unit[2.21]{MHz}$ ($\delta \phi \approx 0$, 7 Points, see Extended Data Fig.~\ref{fig:trapfreq2}) are indicated by blue circles and red diamonds, respectively. The solid line is a fit to the combined data set resulting in $f_0=\unit[1067.74789(40)]{THz}$. The error bars correspond to the errors shown in Fig.~\ref{fig:largefigure} and Extended Data Fig.~\ref{fig:trapfreq2} ($68~\%$ confidence intervals).
}
\label{fig:combineddata}
\end{figure}

The dependence of $\RD$ on the detuning $\DMgH$ from a molecular resonance allows us to perform spectroscopy, since forces arising from coupling to other rotational states are negligible over a large range of detunings (Fig.~\ref{fig:couplingstrength}\textsf{b}).
For this, we have combined data taken for different detunings $\DMgH$ and phases $\delta \phi$ by inverting equation (\ref{eq:signalstrength}) numerically and accounting for the reduced contrast (see Extended Data Fig.~\ref{fig:LatticeRabiFlops}).
The measured Rabi frequency ratios $\RD$ for the \molres transition are summarized in Fig.~\ref{fig:combineddata} together
with a fit to the wings of a simplified model $R_\mathrm{fit}(f)=A/|f-f_0|$ with the free parameters $A$ and $f_0$, where we assumed a detuning-independent background coupling to the far detuned transition in the \Mgtf ion (Fig.~\ref{fig:couplingstrength}\textsf{c}). We determine the transition frequency to be $f_0=\unit[1067.74789(40)]{THz}$ ($\unit[68]{\%}$ confidence interval) which is in agreement with a previous measurement of $f_0=\unit[1067.7473(15)]{THz}$\cite{balfour_rotational_1972}. The fitted characteristic detuning for the Rabi frequency ratio is $A=\unit[45.98(73)]{GHz}$ ($\unit[68]{\%}$ confidence interval).

We estimate that non-destructive detection of the internal state of a molecular ion with near 100~\% efficiency could take less than 10 ms with a larger detuning of $\DMgH\sim\unit[100]{GHz}$ and improvements in the atomic ion's state preparation and manipulation (see Methods). This represents a significant advance over previous destructive detection techniques requiring frequent reloading.
The technique will unfold its full potential in high precision spectroscopy of narrow transitions using an independent spectroscopy laser. However, while in the present work black-body radiation probabilistically populates the detected state, precision spectroscopy will require efficient state preparation schemes\cite{leibfried_quantum_2012, ding_quantum_2012} or ro-vibrational cooling techniques\cite{hansen_efficient_2014, lien_broadband_2014, rellergert_evidence_2013, schneider_all-optical_2010, staanum_rotational_2010}. A combination of these powerful tools will enable the realization of optical clocks based on molecular ions approaching the $10^{-18}$ level\cite{schiller_simplest_2014}, where the underlying clock transitions or a combination of transitions can be sensitive to variations of fundamental constants\cite{schiller_tests_2005}, an electron electric dipole moment (eEDM)\cite{loh_precision_2013} or parity violation in chiral molecules.

\paragraph{Acknowledgements}
We acknowledge the support of DFG through QUEST and Grant SCHM2678/3-1. This work was financially supported by the State of Lower-Saxony, Hannover, Germany. Y.W. acknowledges support from the Braunschweig International Graduate School of Metrology (B-IGSM). We thank E.~Tiemann, H.~Kn\"ockel, O.~Dulieu and I.D.~Leroux for stimulating discussions, M.~Drewsen and O.~Dulieu for providing the transition matrix elements for \MgH, and E.~Tiemann, B.~Hemmerling, and I.D.~Leroux for critical reading of the manuscript.
\paragraph{Author Contributions}
P.O.S. conceived and supervised the experiment. F.W. developed the readout algorithm. The measurements were carried out by F.W., J.C.H. and C.S.. Y.W. performed the simulations and the calculation of the lattice coupling strength. F.W. and P.O.S. wrote the main part of the manuscript. Y.W. and F.G. built essential parts of the experiment. All authors contributed to the discussions of the results and manuscript.
 \paragraph{Author Informations} The authors declare that they have no
 competing financial interests. Correspondence and requests for materials
should be addressed to P.O.S. \\(email: piet.schmidt@quantummetrology.de).

\begin{methods}
\renewcommand{\figurename}{Extended Data Figure}
\subsection{Experimental setup}
The experiments were performed in a linear Paul trap, with typical axial trap frequency of $\unit[2.21]{MHz}$ for $^{25}$Mg$^+$. We use a hyperfine qubit consisting of the states  $\downkmg=\ket{F=3,m_F=3}$ and $\upkmg=\ket{F=2,m_F=2}$ of the $^{2}$S$_{1/2}$ electronic ground state.
The Raman laser system for coherent manipulation of the $^{25}$Mg$^+$ qubit via the $^{2}$P$_{3/2}$  Level is based on a frequency-quadrupled fiber laser\citesec{hemmerling_single_2011}. 
The detuning from the atomic resonance is fixed at $\unit[9.2]{GHz}$, limiting ground state cooling efficiency\cite{wan_efficient_2015} and qubit operations due to spontaneous Raman scattering\citesec{ozeri_hyperfine_2005}. The same laser is also used for internal state discrimination by applying an optical sideband tuned near resonance of the $\downkmg~\rightarrow\ket{^{2}\mathrm{P}_{3/2},F=4, m_F=4}$ component of the D2 transition.
A similar laser system addressing the $^{2}$S$_{1/2}~\rightarrow~^{2}$P$_{1/2}$ (D1) transition is used for optical repumping and quenching during sideband cooling\cite{wan_efficient_2015}.

The light fields creating the optical lattice for molecular state detection and spectroscopy are provided by a frequency-doubled dye laser whose frequency is tuned close to the \molres transition of \MgH with a variable detuning $|\DMgH|\lesssim\unit[100]{GHz}$ from the molecular transition and detuned by $\DMgH\approx \unit[1.5]{THz}$ from the atomic D1 transition of \Mgtf. Two counterpropagating laser beams derived from this laser are linearly polarized in the horizontal plane with a relative detuning $\delta$. They are aligned along the trap axis forming an angle of $\pi/4$ with the atomic quantization axis given by a static applied magnetic field of $\sim\unit[0.58]{mT}$. The detuning between the two laser fields results in a moving optical lattice, providing a temporally and spatially varying force on the two ions with a relative phase $\delta\phi$, that depends on the distance between the two ions. Two optical fibers suitable for UV single mode transmission\citesec{gebert_damage-free_2014}
are used to reduce the effect of beam pointing fluctuations on the coupling strength and aid in the initial alignment of the beams onto the molecular ion. The frequency of the spectroscopy light is monitored by a commercial wavemeter (High Finesse WS/7). We assign an uncertainty of less than \unit[100]{MHz} to the measured value.

To reduce quantum projection noise due to imperfect single qubit operations, we average over 30 measurement cycles to get a single measurement point. This procedure takes $\unit[280]{ms}$. If at least three consecutive points are significantly above the threshold (1.5$\sigma$), we assign it to a positive detection signal from the molecule. For the determination of the signal height, the first and last measurement point were only taken into account if they were at least 2$\sigma$ above the background. This procedure improves the signal quality, since data points are removed which contain only a fraction of positive detection events during the 30 repetitions.
\subsection{Hamiltonian for lattice coupling}
The Hamiltonian for N ions coupled by a 1D moving optical lattice in the interaction picture can be written as
\begin{equation}
H=\sum_{j}^N \hbar \Omega_j \mathrm{e}^{-\ii(\delta t -\phi_j)}\\
\exp\left[ \sum_{k}^N\ii\eta_{jk}\left(a_k\mathrm{e}^{-\ii\omega_k t}+a_k^\dagger\mathrm{e}^{\ii\omega_k t}\right)\right]+\mathrm{h.c.},
\end{equation}
where $j$ labels the ions and $k$ labels the axial motional modes with frequency $\omega_k$. $\eta_{jk}$ is the Lamb-Dicke parameter and $a_k^\dagger$ is the creation operator. $\Omega_j=\Omega_{1j}\Omega_{2j}/(2\Delta)$ is the two-photon Rabi frequency for the two lattice beams coupling to an electronically excited state with detuning $\Delta$.
Considering the case of two ions  $j \in \{1,2\}$ we have to take two axial motional modes into account ($ k \in \{\mathrm{IP,OP}\}$) denoting the in-phase and out-of-phase motion of the two ions, respectively.
In the following, the relative detuning of the two lattice beams, $\delta$ is chosen to be $\delta=\omega_\mathrm{OP}-\omega_\mathrm{IP}$. After the rotating wave approximation only the second order mixed term in the Lamb-Dicke approximation survives, resulting in the following interaction Hamiltonian, mediating the coupling between the two modes:
\begin{equation}
  H=-\sum_{j=1,2} \hbar \Omega_j \mathrm{e}^{\ii\phi_j}\eta_{j,\mathrm{IP}}\eta_{j,\mathrm{OP}}a_\mathrm{IP}a_\mathrm{OP}^\dagger+\mathrm{h.c.}
\end{equation}
In adding up the forces on the two ions we have to consider the different effect on both modes and take the signs in the matrix for transformation between normal modes and laboratory frame into account\citesec{james_quantum_1998}. We find for two ions with equal mass the Lamb-Dicke factors
\begin{eqnarray}
\eta_{1,\mathrm{IP}}&=&\eta_{2,\mathrm{IP}}\equiv\eta_{\mathrm{IP}} ,\\
\eta_{1,\mathrm{OP}}&=&-\eta_{2,\mathrm{OP}}\equiv\eta_{\mathrm{OP}}.
\end{eqnarray}
and the Hamiltonian reads
\begin{eqnarray}
H= -\hbar\eta_{\mathrm{IP}}\eta_{\mathrm{OP}}a_\mathrm{IP}a_\mathrm{OP}^\dagger\left[\Omega_1 \mathrm{e}^{\ii\phi_1}-\Omega_2 \mathrm{e}^{\ii\phi_2}\right]+\mathrm{h.c.}.
\end{eqnarray}
It is convenient to define
\begin{eqnarray}
\delta \phi &\equiv&\phi_1-\phi_2,\\
\Phi&\equiv&\phi_1+\phi_2.
\end{eqnarray}
The Hamiltonian can then be written as
\begin{equation}
H=-\mathrm{e}^{\ii\Phi/2}\hbar\eta_{\mathrm{IP}}\eta_{\mathrm{OP}}a_\mathrm{IP}a_\mathrm{OP}^\dagger\left[\Omega_1 \mathrm{e}^{\ii\delta\phi/2}-\Omega_2 \mathrm{e}^{-\ii\delta\phi/2}\right]+\mathrm{h.c.}.
\end{equation}
We introduce the modified Rabi frequency
\begin{eqnarray}
\Omega=\Omega_\mathrm{eff}\mathrm{e}^{\ii\Phi_\mathrm{eff}}\equiv -\mathrm{e}^{\ii\Phi/2}\left[\Omega_1 \mathrm{e}^{\ii\delta\phi/2}-\Omega_2 \mathrm{e}^{-\ii\delta\phi/2}\right],
\end{eqnarray}
with amplitude
\begin{equation}
\Omega_\mathrm{eff}=\Omega_2\sqrt{1-2R\cos(\delta\phi)+R^2},
\end{equation}
where $R=\Omega_1/\Omega_2$ is the ratio between the Rabi frequencies and $\Phi_\mathrm{eff}$ is an effective phase. Identifying $\Omega_1=\Omega_\mathrm{MgH}$ and $\Omega_2=\Omega_\mathrm{Mg}$ with the Raman Rabi frequencies for the molecular and atomic ion, respectively, we arrive at Eq.~(2) of the main text.
For very large detunings from either the atomic or the molecular ion, the effective Rabi frequency reduces to $\Omega_{\mathrm{MgH}}$ and $\Omega_{\mathrm{Mg}}$, respectively.
The final interaction Hamiltonian is then given by
\begin{equation}
H=\hbar\Omega_\mathrm{eff}\eta_\mathrm{IP}\eta_\mathrm{OP}\left(e^{\ii\Phi_\mathrm{eff}}a_\mathrm{IP}a_\mathrm{OP}^{\dagger}+e^{-\ii\Phi_\mathrm{eff}}a_\mathrm{IP}^\dagger a_\mathrm{OP}\right),
\label{eq:Hint}
\end{equation}
which is a Hamiltonian that describes a system consisting of $n+1$ coupled levels, where $n$ is the number of initial excitations. In the case $n=1$ we get a two-level system consisting of the states $\ket{\downarrow}_\mathrm{m}\equiv \ket{1}_\mathrm{IP}\ket{0}_\mathrm{OP}$, $\ket{\uparrow}_\mathrm{m}\equiv\ket{0}_\mathrm{IP}\ket{1}_\mathrm{OP} $ which are coupled with the Rabi frequency $\Omega_\mathrm{eff}$. The action of this Hamiltonian can be described as Rabi flopping between motional states. Results from an experimental implementation are shown in Extended Data Fig.~\ref{fig:LatticeRabiFlops} for a two-ion crystal consisting of a $^{25}$Mg$^+$ ion and a molecular ion in a rotational state not addressed by the lattice.The reduced initial contrast and its decay are attributed to imperfect ground state cooling, off-resonant scattering of the Raman beams interacting with the atom, and dephasing of the lattice beams. Starting from both axial modes near the ground state, a quantum of motion is added using a BSB $\pi$-pulse. The lattice coupling is then switched on for a variable time and the motional information is mapped back onto the electronic qubit by a second BSB $\pi$-pulse (see Extended Data Fig.~\ref{fig:fullsequence}). In such a sequence, we are not sensitive to the phase factors $e^{\pm\ii\Phi_\mathrm{eff}}$, which can therefore be neglected in the Hamiltonian, resulting in Eq.(\ref{eq:Hamiltonian}) of the main text.
Note that an additional phase of $\pi$ is introduced if the sign of the detuning differs for the atomic and molecular resonances, so the relative phase is given by
\begin{equation}
\delta\phi=
 \begin{cases}
 2\pi d \Delta k & \text{(same sign)}\\
 2\pi( d \Delta k+ \half)  & \text{(opposite sign),}
 \end{cases}
\end{equation}
where $d$ is the distance between ions and $\Delta k=k_1-k_2\approx 2k_1$ is the difference between the wavenumbers of the two lattice beams.

\begin{figure}[htb]
\centering
\includegraphics[width=89mm]{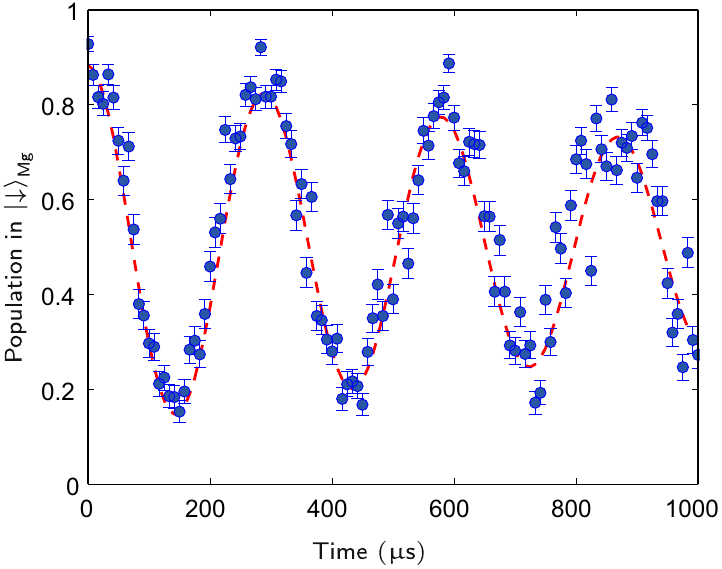}
\caption{\textbf{Rabi flopping between motional states.}
Implementation of the sequence shown in Fig.~\ref{fig:sequence} for $\Omega_{\mathrm{MgH}}=0$. The duration of the applied optical lattice is varied to induce Rabi flopping between the motional qubit states \downkm and \upkm. The error bars show the $95\%$-confidence interval of the photon distribution fit\cite{hemmerling_novel_2012}.
}
\label{fig:LatticeRabiFlops}
\end{figure}
\subsection{Coupling of the optical lattice to the molecular ion }
The coupling strength of the lattice laser to the molecular ion can be described by the transition matrix element between two states $\Ket{\xi^\dprime \Lambda^\dprime \nu^\dprime J^\dprime m_J^\dprime}$ and $\Ket{\xi^\prime \Lambda^\prime \nu^\prime J^\prime m_J^\prime}$ with $\Lambda$ as the projection of the angular momentum on the molecular axis, which is $\Lambda^\prime = \Lambda^\dprime
= 0$ for both involved states, since we have at a $\Sigma\rightarrow\Sigma$  transition without electron angular momentum.
The electronic part of the wavefunctions is given by $\xi^\prime=\mathrm{A}$ and $\xi^\dprime=\mathrm{X}$ in spectroscopic notation. The vibrational and rotational quantum numbers are given by $\nu$ and $J$, respectively, and $m_J$ is the projection of $J$ onto the quantization axis. Note that the quantization axis for the molecule is given by the electric field of the lattice light, since the AC Stark shift causes the dominant level splitting.
The transition dipole matrix element is usually approximated by 
\begin{multline}
   \Bra{\xi^\prime \nu^\prime J^\prime m_J^\prime \Lambda^\prime} d_q \Ket{\xi^\dprime \nu^\dprime J^\dprime m_J^\dprime \Lambda^\dprime}=\\
   \mu_\mathrm{XA}\sqrt{S^\mathrm{FC}(\nu^\prime,\nu^\dprime)S^\mathrm{HL}(J',\Lambda^\prime;J^\dprime,\Lambda^\dprime)S^\mathrm{geom}(J',m_J^\prime;J^\dprime,m_J^\dprime)} \,,
   \label{eq:dme}
\end{multline}
where $\mu_\mathrm{XA}$ is the electronic transition dipole moment.
The Franck-Condon factor describing the overlap of the wave functions of the two
vibrational ground states can be calculated from the data given in Aymar et al.\citesec{aymar_electronic_2009}
\begin{equation}
    S^\mathrm{FC}(\nu^\prime=0, \nu^\dprime=0) = \left[\int\psi_{\nu^\prime=0}^*
    \psi_{\nu^\dprime=0} \dif{r} \right]^2 \approx 0.0919.
\end{equation}
The H\"onl-London factor describing the dependence of the  transition strength on the rotational quantum numbers is
given by Hansson and Watson\cite{hansson_comment_2005}
\begin{equation}
    S^\mathrm{HL}(J^\prime,\Lambda^\prime=0; J^\dprime,\Lambda^\dprime=0)  =
    \max(J^\prime, J^\dprime).
\end{equation}
The geometric factor describing the addition of total angular momentum of the
molecule and the angular momentum of the photon is given by
\begin{equation}
    S_{m_J^\prime q}^\mathrm{geom}(J^\prime m_J^\prime; J^\dprime m_J^\dprime)
    = \left|\epsilon^q \left(\begin{matrix}J^\prime & 1 & J^\dprime \\
              -m_J^\prime & q & m_J^\dprime
    \end{matrix}\right)_{3j} \right|^2,
\end{equation}
with the Wigner 3j-symbol and the polarization components $\epsilon^q$. Since the quantization axis is given by the lattice, it is convenient to calculate the coupling in a basis parallel to the lattice electric field, where the polarization components are $\epsilon^{-1}=\epsilon^1=0$ and $\epsilon^0=1$.

The Rabi frequency from the interaction of the optical lattice with the molecular ion (Eq.~\ref{eq:Hint}) starting from a specific rotational state $\Ket{\xi^\dprime \nu^\dprime
J^\dprime m_J^\dprime}$ is given by\cite{bergmann_coherent_2011}
\begin{eqnarray}
\Omega_\mathrm{MgH}& =& \frac{E_\mathrm{1}E_\mathrm{2}}{2\hbar^2}
\sum_{m_J^\prime}\sum_q \frac{|\epsilon^q \Bra{\xi^\prime \nu^\prime
J^\prime m_J^\prime}d_q
 \Ket{\xi^\dprime \nu^\dprime J^\dprime m_J^\dprime}|^2}{\DMgH} \\
 & =&\frac{E_\mathrm{1}E_\mathrm{2}}{2\hbar^2}
 \frac{\mu_\mathrm{XA}^2}{\DMgH } S^\mathrm{FC}S^\mathrm{HL}
 \sum_{m_J^\prime}\sum_q S_{m_J^\prime q}^\mathrm{geom} \\
& = & \frac{E_\mathrm{1}E_\mathrm{2}}{2\hbar^2}
 \frac{\mu_\mathrm{XA}^2}{\DMgH } S^\mathrm{FC}S^\mathrm{HL} S^\mathrm{geom},
 \label{eq:couplStrength}
\end{eqnarray}
where $E_1$ and $E_2$ are the electric field amplitudes of the two counterpropagating laser beams forming the optical lattice and $\DMgH$ is the detuning from the molecular resonance.
Using a similar approach for the atomic ion, we get a Rabi frequency ratio that is independent of the applied electric field. From the well-known transition strengths in \Mgtf and the fitted value for $A=45.98(73)$~GHz (68~\% confidence interval), we extract a dipole matrix element for the molecular transition of $\mu_{XA}^\mathrm{exp}=1.2510(99)~ea_0$, which is smaller than the matrix element $\mu_\mathrm{XA}^\mathrm{th} = 1.779\,ea_\mathrm{0}$ extracted from the theoretical data presented in Aymar et al.\citesec{aymar_electronic_2009}.
We use the experimental dipole matrix element for the curves in Fig.~\ref{fig:largefigure} and Extended Data Fig.~\ref{fig:trapfreq2}.
\subsection{Scattering rate and estimated single-cycle detection efficiency}
We assume that scattering of a single photon on the optical transition of the molecular ion will change its ro-vibrational state, due to the non-vanishing Franck-Condon factors between different vibrational states. Therefore, a small scattering rate is desired. In the following we use a simplified model in which the molecular ion is treated as a two-level system to estimate the order of magnitude for the spontaneous emission rate of the molecular ion subject to the ODF.
The number of scattered photons from a two-level system with excited state linewidth $\Gamma_0$ interacting for a time $t_\pi$ with an optical lattice formed by two laser beams of equal Rabi frequency $\Omega$ and detuned by $\DMg$ from resonance is given by
\begin{eqnarray}
\Gamma_\mathrm{sc} t_\pi = \frac{\Gamma_0\Omega^2}{2\DMgH^2} t_\pi.
\label{eq:Ramansc}
\end{eqnarray}
For the implemented quantum algorithm the time $t_\pi$ is chosen such that the coupling to the atomic ion drives a $\pi$ pulse between the two motional states, $t_\pi=\pi/(\Omega_\mathrm{Mg}\eta_\mathrm{IP}\eta_\mathrm{OP})$. Using $R=\Omega_\mathrm{MgH}/\Omega_\mathrm{Mg}$ and $\Omega_\mathrm{MgH}=\Omega^2/(2\DMgH)$, we obtain
\begin{eqnarray}
\Gamma_\mathrm{sc} t_\pi & =& \frac{\Gamma_0 \pi R}{\DMgH \eta_\mathrm{IP} \eta_\mathrm{OP}},
\end{eqnarray}
where $\Gamma_\mathrm{0}$ for the molecular ion is given by
\begin{equation}
    \Gamma_0 = \frac{\omega_\mathrm{0}^3}{3\pi \epsilon_\mathrm{0}\hbar,
c^3} \mu_\mathrm{XA}^2.
\end{equation}
with the scaled transition dipole moment $\mu_\mathrm{XA}$ given above.
For Poissonian photon distribution the probability to scatter \textbf{no} photon follows
\begin{equation}
P_\mathrm{scatt}(0)=\mathrm{e}^{-\Gamma_\mathrm{sc}t_\pi}.
\end{equation}
For a detuning of $\approx\unit[25]{GHz}$ we expect to scatter on average 0.04 photons, which corresponds to a fundamentally limited single cycle detection efficiency of $E_\mathrm{SS}=P_\mathrm{scatt}(0)\approx \unit[96]{\%}$. For an increased detuning of $\unit[-100]{GHz}$ 
the number of scattered photons on average would be reduced to $0.0026$ for a $\pi$-pulse with the optical lattice. However, to get a sufficiently large signal contrast for single-cycle detection it would be necessary to perform a $2\pi$-pulse with the lattice, which leads to a single-cycle detection efficiency of $\unit[99.5]{\%}$.

The number of positive detection events $N_\mathrm{ev}$ for a total number $N_\mathrm{total}$ of trials can be estimated from
\begin{equation}
N_\mathrm{ev}=(E_\mathrm{SS})^{3N_\mathrm{cycle}}P_\mathrm{therm}P_\mathrm{t\leq3} N_\mathrm{total},
\end{equation}
where $P_\mathrm{therm}$ is the probability for the molecule to be in the probed state for a thermal distribution at 300~K and $P_\mathrm{t>3}$ is the probability for the molecular ion to stay at least for the time of three measurement points in the probed state, which was a chosen prerequisite for positive detection in the experiment.
In our case, we have  $N_\mathrm{total}\approx 2\times 10^5$, $P_\mathrm{therm}=0.08$, $P_\mathrm{t\leq3}=0.8$ (see Monte Carlo simulation in Extended Data Fig.~\ref{fig:MCSimu}) and $N_\mathrm{cycle}=30$. For a single shot detection efficiency on the order of $E_\mathrm{SS}\approx\unit[95]{\%}$, this results in an expected number of points belonging to a detection signal of $N_\mathrm{ev}\approx 130$, which is in agreement with the observed number of points $N_\mathrm{ev}^\mathrm{exp}\approx 100$ assigned to the 18 positive detection events.
\subsection{Uncertainties of the fitted parameters}
We extracted the center frequency and the line strength from a weighted non-linear least square fit. The assigned uncertainties correspond to the $\unit[68]{\%}$ confidence bounds $C$, that have been calculated assuming a normal distribution of errors and according to $C=b\pm t\sqrt{S}$, where $b$ is the fitted coefficient, $t$ is the inverse of Student's cumulative distribution function for the required confidence and $S=(X^\mathrm{T}X)^{-1}s^2$, with the mean squared error $s^2$ and the Jacobian $X$ of the fitted values with respect to the coefficients.
\begin{figure}
\centering
\includegraphics[width=89mm]{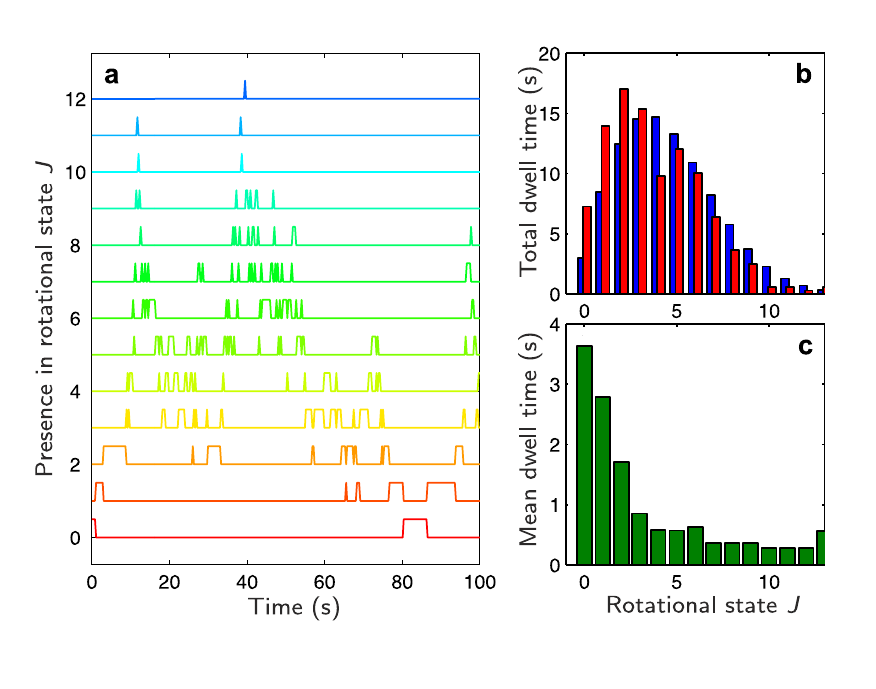}
\caption{\textbf{Single trajectory from a Monte Carlo simulation of molecular dynamics.} (\textsf{a}) The ion initially prepared in the rotational ground state is transfered to higher rotational states due to coupling to BBR at \unit[300]{K}. (\textsf{b}) The probability to find the ion in a certain rotational state in the simulation (red bars) follows a thermal distribution. The blue bars are calculated values from a master equation approach. The deviation is due to the finite time interval of the Monte Carlo wave function simulation. (\textsf{c}) The dwell time decreases for higher rotational states.}
\label{fig:MCSimu}
\end{figure}
\begin{figure}
\centering\includegraphics[width=89mm]{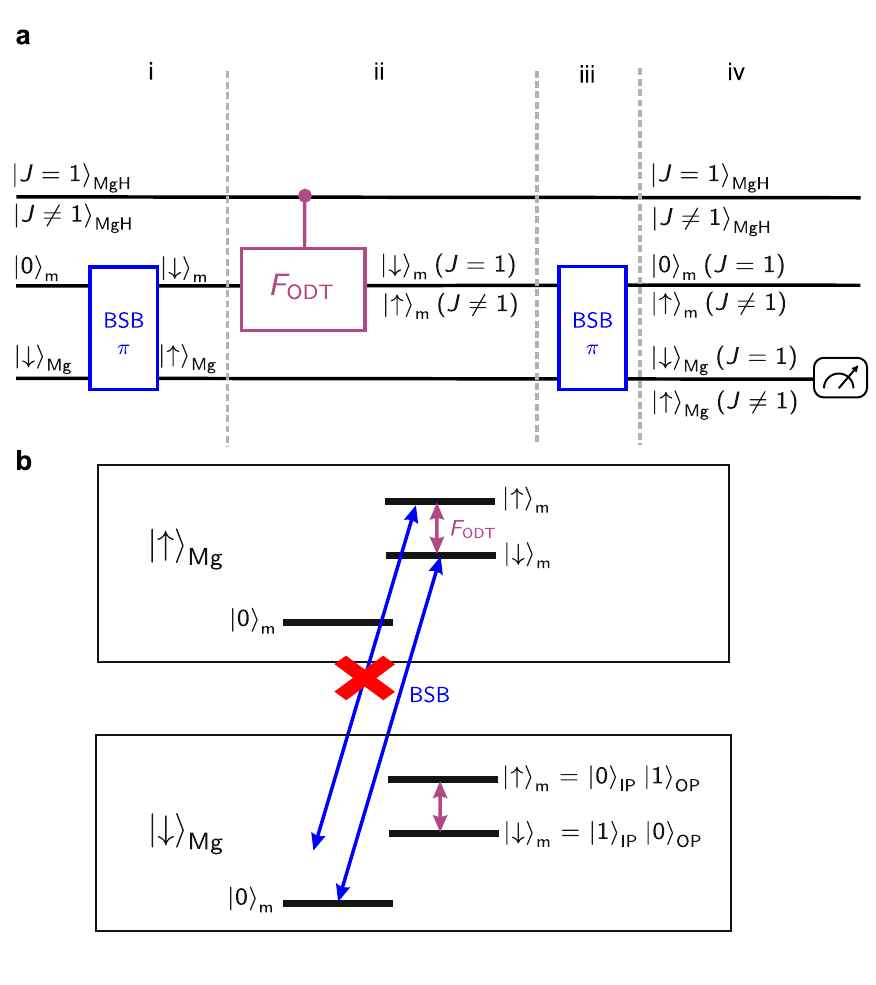}
\caption{\textbf{Full experimental sequence.} 
(\textsf{a}) Circuit description of the sequence. \textsf{(i)} A BSB $\pi$-pulse initializes the atom in the state $\upkmg$ and the motional state in the qubit state $\downkm$. (\textsf{ii}) The ODF rotates the motional qubit controlled by the internal state of the molecule (see Fig.~\ref{fig:sequence}). (\textsf{iii}) A second BSB $\pi$-pulse maps the motional state (that contains the information about the internal state of the molecule) to the atomic qubit. (\textsf{iv}) The atomic qubit is read out by state dependent fluorescence.
(\textsf{b}) Pictorial representation of the laser couplings in a simplified level scheme.}
\label{fig:fullsequence}
\end{figure}
\begin{figure}
\centering\includegraphics[width=89mm]{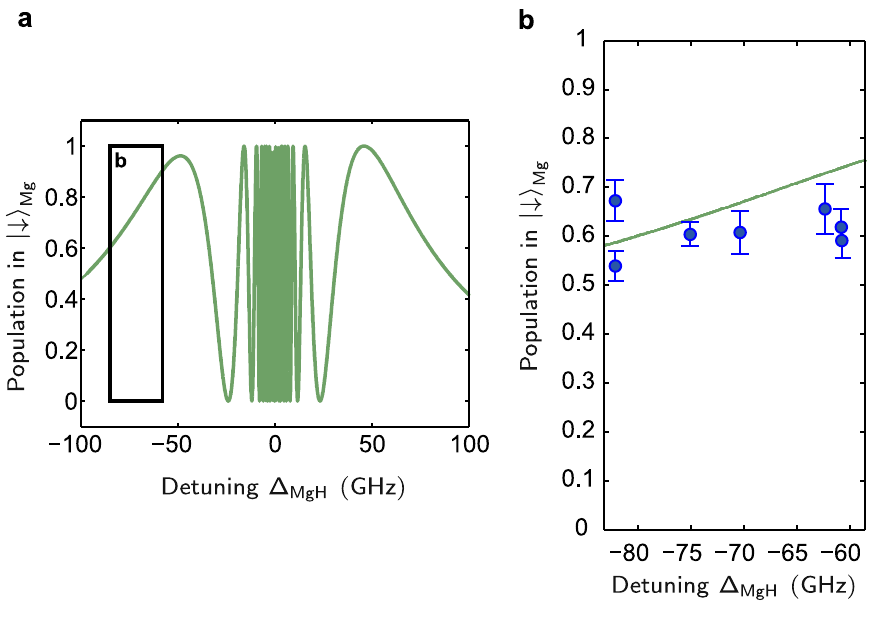}
\caption{\textbf{Raw data for \boldmath$\delta\phi\approx0$.}  (\textsf{a}) Theoretically predicted signal for $\delta\phi\approx0$, corresponding to $\omega_\mathrm{IP}\approx\unit[2.21]{MHz}$. (\textsf{b}) Zoom into the region shown in \textsf{a} with the measured 7 data points included as red diamonds in Fig.~\ref{fig:combineddata}. The error bars indicate the $\unit[68]{\%}$ confidence interval of the photon distribution fit\cite{hemmerling_novel_2012}. 
}
\label{fig:trapfreq2}
\end{figure}

\end{methods}

\end{document}